# *Spin Seebeck Effect in Correlated Antiferromagnetic $V_2O_3$*

Renjie Luo[1], Tanner J. Legvold[1,†], Gage Eichman[1], Henry Navarro[2,3], Ali C. Basaran[2,4], Erbin Qiu[2], Ivan K. Schuller[2], Douglas Natelson[1,5,6,7,*]

[1]*Department of Physics and Astronomy, Rice University, Houston TX 77005, USA*

[2]*Department of Physics and Center for Advanced Nanoscience, University of California-San Diego, La Jolla, CA 92093, USA*

[3]*Department of Physics, Andrews University, Berrien Springs, MI 49104, USA*

[4]*General Atomics, San Diego, CA 92121, USA*

[5]*Department of Electrical and Computer Engineering and Department of Materials Science and NanoEngineering, Rice University, Houston TX 77005, USA*

[6]*Rice Center for Quantum Materials, Smalley Curl Institute, Rice University, Houston, TX 77005, USA*

[7]*Rice Advanced Materials Institute, Rice University, Houston, TX 77005, USA*

**Abstract**

The spin Seebeck effect is useful for probing the spin correlations and magnetic order in magnetic insulators. Here, we report a strong longitudinal spin Seebeck effect (LSSE) in antiferromagnetic $V_2O_3$ thin films. The LSSE response at cryogenic temperatures increases as a function of the external magnetic field until it approaches saturation. The response at given power and field exhibits a non-monotonic temperature dependence, with a pronounced peak that shifts toward higher temperatures as the field increases. Furthermore, the magnitude of the LSSE signal decreases consistently with increasing thickness, implying that the bulk SSE dominates any interfacial contribution. This negative correlation between the SSE and the thickness implies that the magnon energy relaxation length in $V_2O_3$ is shorter than the thickness of our thinnest film, 50 nm, consistent with the strong spin-lattice coupling in this material.

[*] natelson@rice.edu

[†] deceased.

The spin Seebeck effect (SSE) [1–4] generates spin current between a magnetic insulator (MI) and a nonmagnetic metal under heat, and it has proven to be a powerful tool to study spin correlations and magnetic order in insulators [4]. Extensive investigations have explored the SSE across diverse magnetic systems– including ferrimagnets [5–8], ferromagnets [9], antiferromagnets [10–23], quantum spin systems [24-28], and paramagnets [20, 29, 30]. Among these, antiferromagnetic insulators (AFIs) are of great interest because they have no net magnetization to couple with external fields and show terahertz frequency magnetization dynamics, making them promising materials for spintronic applications [31]. However, SSE in AFIs is comparatively less understood due to the presence of multiple magnon branches, spin-flop transitions, and complex responses [10-23].

Vanadium sesquioxide ($V_2O_3$), an archetypal material with strong electronic correlations, undergoes a high-temperature rhombohedral paramagnetic metal to a low-temperature monoclinic antiferromagnetic insulator transition around 160 K. Neutron diffraction reveals a magnetic order where V moments that are ferromagnetically coupled in monoclinic (010) layers tilted away from the hexagonal c-axis by 71° [32], with a reversal between adjacent layers (Fig. 1a). The ordered moment is (1.2 ± 0.1) $\mu_B$ per V atom, and the spin excitation gap is 4.75 meV (1.15 THz, or 55 K) [33]. Resonant x-ray scattering experiments at the vanadium K edge found additional Bragg peaks due to the long-range order of 3$d$ orbital occupancy and demonstrated the existence of orbital ordering in $V_2O_3$, which could account for its complex magnetic structure [34]. While $V_2O_3$ has often been discussed as an example of a Mott insulator, the situation is quite subtle [35]. The nature of the metal-insulator phase transition in $V_2O_3$ is complicated, with strong couplings between spin, lattice, and charge degrees of freedom [36]. The antiferromagnetic (AFM) order in $V_2O_3$, as evidenced in Ni/$V_2O_3$/permalloy spin-valve devices, induces unexpected temperature dependence across the $V_2O_3$ metal-insulator transition [37]. The AFM order can also significantly enhance the exchange bias of an attached ferromagnetic permalloy layer, which appears at the onset of the transition temperature [38]. However, to date, spin calorimetric measurements have not been applied to $V_2O_3$.

In this work, we have investigated a local spin Seebeck effect (LSSE) in antiferromagnetic $V_2O_3$ thin films as a function of temperature, magnetic field, and film thickness using Pt electrodes. At a fixed heater power and low temperature, the LSSE increases with field until it approaches saturation. At fixed heater power and field, the LSSE has a non-monotonic temperature dependence, with a pronounced peak at low temperatures that shifts toward higher temperatures with increasing field. In contrast to the LSSE observed in paramagnetic $VO_2$ [30], the SSE signal in AFM $V_2O_3$ decreases monotonically as a function of film thickness, implying that the measured response stems from the bulk. These results imply that the thermal relaxation length between magnons and phonons in $V_2O_3$ is less than 50 nm, consistent with the wealth of experimental evidence for strong spin-lattice interactions in this material.

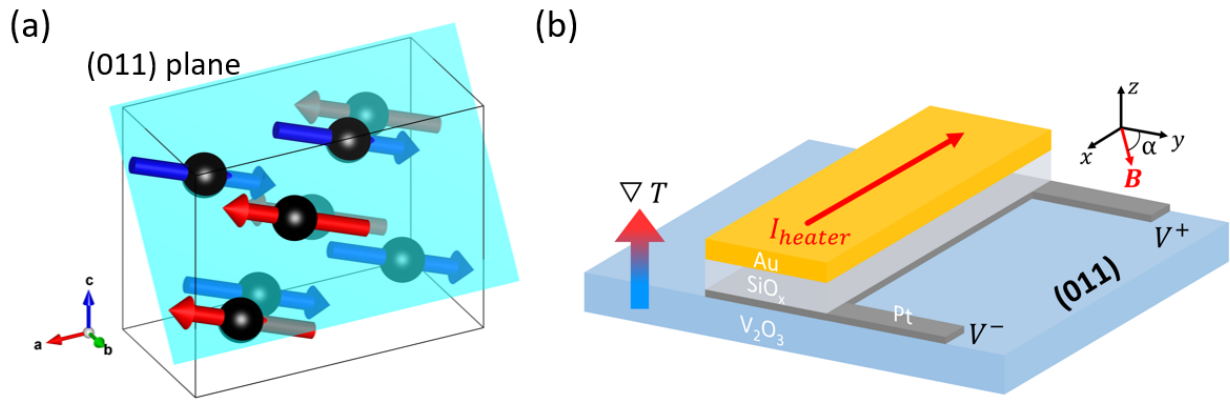

FIG 1. (color online) (a) Crystal structure of $V_2O_3$ in the low temperature, insulating monoclinic phase. The black spheres represent V atoms, with red and blue arrows showing the spin direction. The moments of V atoms are aligned ferromagnetically on the (010) plane and reverse between the neighboring layers. The cyan plane indicates the monoclinic (011)-oriented surface, which is oriented along $z$ in the devices examined. Oxygen atoms are omitted for clarity. The crystal structure is generated by VESTA [39]. (b) Schematic of local spin Seebeck measurement. An AC heater current produces an oscillating $z$-directed temperature gradient. A vertical ($z$-directed) thermal spin current with the $y$ component of the magnetization of the $V_2O_3$ would produce an ISH voltage along the $x$-directed strong spin-orbit metal wire.

We grew $V_2O_3$ thin films (50, 100, 250, and 400 nm) epitaxially on r-cut sapphire substrates via RF magnetron sputtering. Due to the lattice match, the $V_2O_3$ thin film favors a (011)-oriented texture (in the low temperature monoclinic phase), where the moments of V are almost parallel to the surface, and the surface is nominally magnetically compensated. X-ray diffraction (XRD) measurements found essentially identical diffraction peaks for the different film thicknesses reflecting the high temperature (012)/low temperature (011) orientation. The observed XRD peak widths and the smooth, featureless background confirm that the films remain single-phase, textured, and structurally coherent across the thickness range. A slight increase in out-of-plane spacing with thickness indicates partial relaxation of substrate-induced strain. The Pt (or W) wire, 800 μm long, 10 μm wide, and 10 nm thick, was directly deposited on the $V_2O_3$ film surface using standard photolithography, magnetron sputtering, and liftoff process, with wire orientations parallel to the edge of the substrate. Because of the texture of the film, the wires are averaging over all in-plane crystallographic orientations. A $SiO_x$ layer with a thickness of 100 nm and an Au heater wire (1300 μm long, 10 μm wide, 50 nm thick) were then fabricated above the Pt (or W) wire by photolithography, e-beam evaporation, and liftoff. The Au heater wire and the Pt detector wire are electrically isolated by the $SiO_x$ layer in between. An AC current at angular frequency $\omega = 2\pi \times (7.7$ Hz$)$ is driven through the Au wire, while the voltage across the Pt (or W) wire is recorded at $2\omega$ using a lock-in amplifier. The $2\omega$ $y$-component is the relevant signal, though both phases of $\omega$ and $2\omega$ voltages were measured. The phase lag was very small ($< 5°$) over the full temperature range of measurements. The measurements are performed as a function of temperature and field in a Quantum Design Physical Property Measurement System (9T-PPMS) equipped with a rotation stage. The films all have the metal-insulator transition at the expected temperature range and are extremely insulating at cryogenic temperatures.

The schematic of the device is given in Fig. 1b. In the on-chip-heating configuration for detecting the LSSE [40], an alternating current at frequency $\omega$ is applied to a heater wire. Local Joule heating generates a temperature gradient along the normal direction of the film surface, oscillating at $2\omega$, which in turn drives a spin angular momentum current. A $2\omega$ transverse voltage along the $x$-direction can be detected at a nearby inverse spin Hall (ISH) detector made from a strong spin-orbit metal (e.g., Pt, W) when the magnetization of the insulator is oriented in the $y$ direction. Thermal modeling of this device geometry [30, 41] shows that thermal transport from the heater into the film is essentially vertical because the film thickness is far smaller than the transverse dimensions of the Pt or Au strips. Prior experiments [41] on $SiO_2$ substrates for much larger heater powers than this work find the typical temperature rise of the Pt in this geometry is

~ 1 K at 1 mW heater power at the lowest temperatures. The absence of saturation in the signal vs. cryostat temperature in the present measurements is consistent with the sample temperature being close to the PPMS recorded temperature.

Magnetic field dependence of the second harmonic voltage $V_{2\omega}$ is shown for a 100 nm thick Pt/V$_2$O$_3$ device when the field is applied in the film plane along the y-axis ($\alpha = 0°$) (Fig. 2a, b). At $T = 2$ K, a pronounced $V_{2\omega}$ signal appears, whose sign changes with respect to the **B** direction, reflecting the symmetry of the ISHE. The signal shows approximately linear dependence near $|\boldsymbol{B}|$ = 0 T, then saturates at high fields at 5 T. A full sweep of the field between ±8 T exhibits no hysteresis. As the temperature increases from 2 to 7 K, the overall magnitude of $V_{2\omega}$ increases, and the saturation region shifts to higher fields (Fig. 2a). However, above 7 K the signal magnitude decreases and saturation behavior disappears (Fig. 2b). The signal vanishes above $T = 50$ K. In Fig. 2c, the $V_{2\omega}$ signal is normalized to its maximum value at 8 T for each temperature, to highlight the nonlinear curvature at $T = 2$ K. As temperature increases, the nonlinearity becomes weaker with the decrease of the zero-field slope. This field dependence of $V_{2\omega}$ closely parallels to uniaxial AFI MnF$_2$ [10], FeF$_2$ [15] with field along the easy axis below the spin-flop fields, and also in insulating SrFeO$_{3-\delta}$ thin films [14]. Recently, a similar field dependence was reported in uniaxial AFI Cr$_2$O$_3$ when a field is applied along its hard-axis and interpreted as paramagnetic SSE from weakly interacting spins near the Cr$_2$O$_3$ surface [23].

As a control experiment, the voltage responses for devices with Pt and W detectors show opposite signs as seen in Fig. 2d. This is consistent with a genuine spin current effect, since Pt and W are known to possess spin Hall angles of opposite signs [42]. Aside from this sign reversal, the field dependence of the voltage in Pt/V$_2$O$_3$ and W/V$_2$O$_3$ is similar, indicating a common origin.

We further confirmed the magnetic origin of the measured signal by rotating the in-plane magnetic field while keeping the field magnitude constant ($|\boldsymbol{B}| = 1$ T, 3 T, 5 T, 7 T) at 3 K. A cosine dependence of the $V_{2\omega}$ signal was observed (Fig. 2e). The SSE voltage reaches its maximum magnitude when the field is perpendicular to the wire (along the y-axis); and the voltage crosses zero when the field aligns along the wire (along the x-axis). The voltage from ISHE is given by $\boldsymbol{E}_{ISHE} \propto \boldsymbol{J}_s \times \boldsymbol{\sigma}$, where $\boldsymbol{J}_s$ is the interfacial spin current density into the Pt and $\boldsymbol{\sigma}$ is the spin-polarization vector. Note that in our local SSE setup, $\boldsymbol{J}_s$ is always directed along the z-axis parallel to the temperature gradient and $\boldsymbol{\sigma}$ aligns with the magnetization transported in the magnetic insulator.

The observed dependence of $V_{2\omega}$ signal on in-plane field orientation is the same as observed in ferrimagnets and paramagnets [5,24]. Multiple mechanisms were suggested to explain this dependence in antiferromagnets. In AFMs, the external magnetic field can produce a net magnetization component along the field direction due to canting [10]. Alternatively, there are two magnon branches that carry opposite senses of angular momentum. The energy gap associated with each branch depends via Zeeman physics on the orientation and magnitude of the magnetic field. It has been suggested in $FeF_2$ [15] reorienting $B$ in the plane modulates the Zeeman splitting of the magnon modes sinusoidally, leading to the same angular dependence of transported spin. The SSE voltage scales linearly with the applied heater power ($P$) at two different $|B|$ = 1 and 3 T applied fields along the y-axis ($\alpha = 0°$) (Fig. 2f).

Another effect with the same field orientation dependence in this geometry is the ordinary Nernst-Ettingshausen response of Pt (W). The Nernst-Ettingshausen effect, related to transverse scattering and its coupling to thermoelectricity, is linear with the applied magnetic field [41], and thus cannot explain the observed nonlinear magnetic field dependence shown in Fig. 2a-c. Furthermore, the previous study with the Pt (W) wire deposited directly on the $SiO_2$/Si substrate using an identical geometry yields a Nernst signal of only ~ 20 nV at 8 T under 1 mW heater power [41], which is two orders of magnitude smaller than the signals measured in Pt/$V_2O_3$ devices.

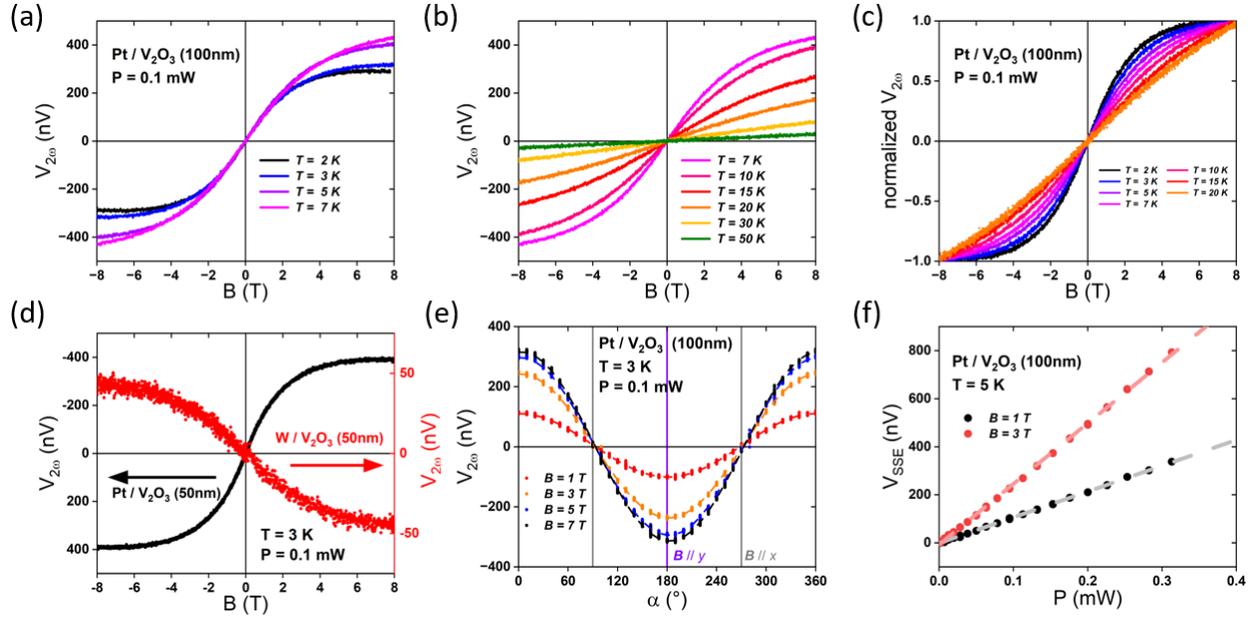

FIG 2. (color online) (a,b) The second harmonic voltage as a function of field ($\alpha = 0°$, $\boldsymbol{B}//y$) at various temperatures for Pt detector wires on 100 nm thick $V_2O_3$. (c) The same dataset in (a,b) with normalization to the maximum value at each temperature. A transition from nonlinearity to linearity can be seen as temperature increases. (d) Comparison between voltage responses of Pt/$V_2O_3$ and W/$V_2O_3$ devices at 3 K with an applied heater power of 0.1 mW. (e) Dependence of the second harmonic signal in Pt wire at 3 K with 0.1 mW heater power on in-plane field orientation $\alpha$, showing expected cosine dependence. (f) Dependence of the spin Seebeck voltage on the heater power at 3 K and $\alpha = 0°$. The dark dots are the measurement data, with corresponding light dash lines showing linear behavior.

To elucidate the mechanism driving the spin Seebeck response in $V_2O_3$, we examined its temperature dependence. Fig. 3a shows the LSSE voltage versus temperature in the same Pt/$V_2O_3$ device at fixed heater power and magnetic field for four applied fields. At each field, the LSSE voltage increases as the temperature decreases, reaches a maximum at a peak temperature $T_{peak}$. After the peak, the signal decreases with further reducing temperature. Such peak behavior is also observed in the ferrimagnet $Y_3Fe_5O_{12}$ (YIG) [43] and AFI $MnF_2$ [10] and $FeF_2$ [15].

In ferrimagnetic YIG, the peak temperature in SSE voltage correlates with the peak temperature of the thermal conductivity of YIG [43], explained by the phonon-drag mechanism. This mechanism involves temperature-gradient-driven phonons that induce the spin current via magnon-phonon interactions. In this case, the phonon-drag model [44] predicts $V_{SSE} \propto \kappa \nabla T = j_q$, where $\kappa$ is the thermal conductivity, $\nabla T$ is the temperature gradient, and $j_q$ is the heat flux through the interface. However, $j_q$ is considered to be fixed in our measurement method, since the heater power is held fixed and the thermal transport in our geometry is essentially all vertical [30, 41].

In AFI MnF$_2$ and FeF$_2$ [10,15], the peak behavior is explained in terms of the antiferromagnetic magnon modes. The uniaxial AFI has two magnon modes (α and β) with opposite chirality. Since β mode lies below α mode under an external field, only the β mode magnons are predominantly populated at low temperatures. As the temperature increases, the β mode magnons population continues to increase until the α mode magnons start to be occupied. Because of the opposite angular momentum between α and β modes, the increasing temperature results in a peak in the net spin current and consequently a peak in SSE voltage.

In V$_2$O$_3$, the peak temperature increases with increasing fields (Fig. 3b), similar to that in MnF$_2$. However, in FeF$_2$, the peak temperature remains constant for all fields, an aspect that is not well understood at present. The qualitative temperature dependence of $V_{SSE}$ and its variation with field (higher field leads to higher peak temperature) are very similar to what is observed in paramagnetic VO$_2$ [30]. We shall discuss this further in the following section.

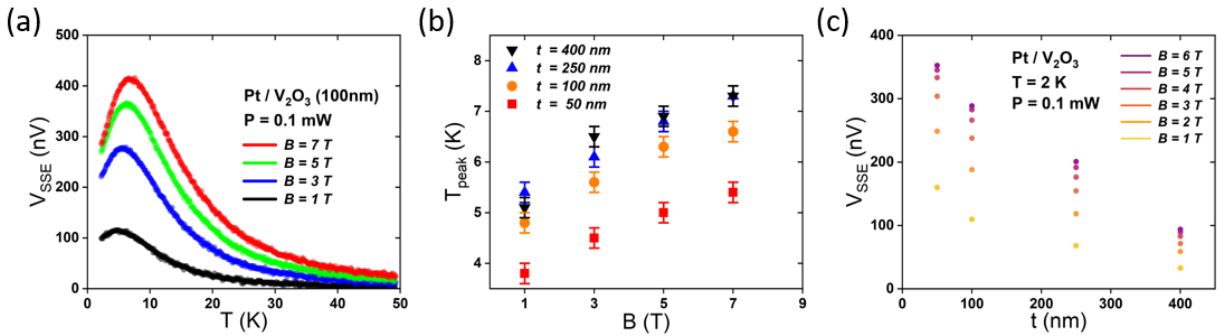

FIG 3. (color online) (a) Temperature dependence of the LSSE voltage, defined as the difference of the second harmonic voltages between at a particular field and 0 T, in the Pt/V$_2$O$_3$ device at constant heater power of 0.1 mW and α = 0°. (b) Field dependence of the peak temperature of the LSSE signal in devices on different thicknesses of V$_2$O$_3$. For each thickness, the peak temperature approximately linearly increases with the field. (c) The thickness dependence of the LSSE voltage at 2 K in Pt/V$_2$O$_3$ devices at different fields. As thickness increases, the magnitude of the LSSE voltage decreases.

The LSSE in an AFI has both a bulk contribution [45,46] from the temperature gradient in the AFI and an interfacial contribution [47–50]. The interfacial contribution comes from the temperature difference between the magnon temperature in the AFI and the electron temperature in the spin-orbit metal. To distinguish between dominating bulk and interfacial contributions, we fabricated devices with identical geometry, following the same fabrication protocols, but varying the thicknesses of the $V_2O_3$ films. It is important to note that all devices exhibit similar magnetic field and temperature dependencies of the LSSE signal, including linearity around zero field, saturation at higher fields, weaker nonlinearity with increasing temperature, peak behavior with temperature, and an increase in peak temperature with increasing field. However, strong thickness dependence appears as the peak temperature at each field shifts upward for thicker films (Fig. 3b) and the magnitude of the LSSE signal decreases consistently as the thickness increases (Fig. 3c). We should consider possible mechanisms in the observed thickness dependence.

The spin Seebeck coefficient [43,51] is defined as $\sigma_{SSE} = (V_{SSE}/l)/(dT/dz)$, where $l$ is the length of the ISH detector, and $dT/dz$ is the temperature gradient across the insulator. At low temperatures and with thin film materials, it is difficult to determine $dT/dz$ experimentally, as this would require measuring the local temperature in the magnetic insulator just below the interface with the ISH detector and just above the interface with the film's underlying substrate. These interfacial local temperatures are generally not accessible, and at low temperatures the situation is further complicated by enhanced thermal boundary resistances due to phonon mismatch. This situation may be mitigated to some degree by working at fixed heater power, and hence approximately fixed vertical heat flux, $j_q$, through the film. One can then know or estimate the film cross-plane thermal conductivity, $\kappa(T)$, or instead work with the spin Seebeck resistivity [52], which is directly proportional to $V_{SSE}$ for fixed sample geometry and heater power.

Guo *et al.* found that in YIG films with thicknesses ranging from 150 nm to 50 μm, the magnitude of the spin Seebeck coefficient $\sigma_{SSE}$ increases gradually with thickness and saturating beyond a critical thickness. The peak temperature shifts to a higher temperature as the film becomes thinner [52]. These observations are explained by the combination of a temperature-dependent magnon diffusion length $\lambda_m \propto T^{-1}$ and the thermal magnon density $\rho_m(T)$. Only magnons excited within a distance $\sim \lambda_m$ from the interface can reach the detector and contribute to the spin current. Thus, the SSE signal scales as $V_{SSE} \propto \rho_m \times \min(\lambda_m, t)$. As the temperature decreases from room temperature, $\lambda_m$ increases and $\rho_m$ decreases. As $\lambda_m$ increases, the SSE signal initially increases, allowing more magnons to reach the interface. However, once $\lambda_m$ reaches the film thickness $t$, further cooling decreases $\rho_m$, causing the SSE signal to drop. This competition leads to a non-monotonic temperature dependence. In thinner films, $\lambda_m$ reaches $t$ at higher temperatures, resulting in the shift of the peak to higher temperatures.

In the Pt/V$_2$O$_3$, vertical thermal conduction at fixed heater power would imply a fixed interfacial temperature difference $T_p(\text{Pt})$-$T_p(\text{V}_2\text{O}_3)$ (for unchanging thermal boundary resistance) and a fixed temperature gradient $\nabla T_p(\text{V}_2\text{O}_3)$ within the film, where $T_p(\text{Pt})$ and $T_p(\text{V}_2\text{O}_3)$ are the phonon temperatures of Pt and V$_2$O$_3$, respectively. If we assume that the spin current is generated from the thermally excited magnons in the bulk V$_2$O$_3$, then the spin Seebeck voltage would be expected to increase as the thickness increases and saturate when the thickness reaches $\lambda_m$ [45,46]. However, instead, we observe a decrease in the signal as the thickness increases. Matching the observation requires the interfacial spin mixing conductance to decrease faster than the thickness factor increases, implying thicker films have more defects [23].

Another possible mechanism for decreasing SSE response with thicker films could arise from a thickness dependent perpendicular thermal conductivity, κ, of V$_2$O$_3$. If κ increases with thickness, the temperature gradient at fixed heater power would be lower for thicker films. Matching the observation requires roughly an inverse relationship between thermal conductivity and thickness. However, there is no reason to expect significant variation of κ with thickness.

A particular explanation for the observed thickness dependence of the SSE signal is consistent with the strong spin-lattice coupling known in V$_2$O$_3$. An alternative approach to quantifying SSE responses uses the spin Seebeck resistivity [52], defined as $R_{\text{SSE}} = (V_{\text{SSE}}/l)/j_q$, where $j_q$ is the heat flux through the insulator. Prakash *et al.* found that the SSE is a non-monotonic function of YIG thickness ranging from 10 nm to 1 um with a peak at ~ 250 nm in Pt/YIG/gadolinium gallium garnet (GGG) stacks, which is explained in terms of the energy-equilibration dynamics of magnons to phonons in the YIG. The magnon energy relaxation length $\lambda_u$ parametrizes the relaxation of the magnon temperature $T_m$ to the phonon temperature $T_p$. When the film thickness is close to $\lambda_u$, thermalization mechanisms between magnons in YIG and phonons in GGG lower the magnon temperature $T_m$ near the YIG/GGG interface, which contribute positively to the temperature imbalance $T_p - T_m$ in YIG. However, when the film thickness is larger than $\lambda_u$, these magnons thermalize with phonons before reaching the Pt/YIG interface, thus this contribution vanishes, leading to the peak behavior in thickness. If we assume the spin current is from the difference between $T_m$ and $T_p$ in the bulk of the $V_2O_3$, then the observed thickness dependence would imply that the decrease in SSE voltage with increasing thickness is a consequence of $\lambda_u$ being smaller than 50 nm, which is the thinnest studied sample here. Further measurements on ultrathin films (i.e., those thinner than 50 nm) and bulk crystals would provide more insight.

We note that our previous studies found that the paramagnetic insulator vanadium dioxide ($VO_2$) supports the paramagnetic SSE at low temperatures [30]. The current investigation finds that AFI $V_2O_3$ and paramagnetic insulator $VO_2$, though magnetically distinct, exhibit SSE signals with the same sign, comparable magnitude under the same heater power with identical geometry, and similar temperature and field dependences. We observed a field-induced suppression of the SSE voltage at low temperature and high field in $VO_2$, whereas $V_2O_3$ does not exhibit this suppression. Still, the similarities in the LSSE response of the two oxides warrant discussion, and it appears that the interface between oxide and Pt could potentially play a role.

Since $VO_2$ and $V_2O_3$ were both grown on r-cut sapphire, the (012) plane of $Al_2O_3$ favors a lattice-matching V atom arrangement analogous to Al atoms on the surface of $Al_2O_3$, ignoring small lattice distortions. The Pt/oxide interface is where the exchange interaction between $d$ electrons of V and $s$ electrons of Pt occurs, inducing the spin current into Pt. We also notice that the SSE in $Cr_2O_3$ when the field is along the hard axis is interpreted as paramagnetic in nature from weakly interacting spins near the $Cr_2O_3$ surface. Thus, it is important to consider whether the SSE signal in $V_2O_3$ can be understood as the paramagnetic SSE.

In a paramagnetic insulator (PI), the interfacial spin current, and thus the SSE voltage, is expected to be roughly proportional to the magnetization of the PI described by a Brillouin function $B_S(x)$ of the spin S, where $x = g\mu_B B/k_B T$ [29,53]. To check this, we examine whether the field dependence of the voltage can be fitted with a Brillouin function. Under this assumption, the SSE voltage can be written as:

$$V(B,T) = C(T)B_S\left(\frac{g\mu_B B}{k_B(T - \Theta_{CW})}\right)$$

where $C(T)$ is temperature-dependent magnitude including thermal conductivity, and $\Theta_{CW}$ is a possible Curie-Weiss temperature. Fig. 4 shows the fitting results with and without $\Theta_{CW}$. The fitting excluding $\Theta_{CW}$ fails to capture the shape of the data accurately, whereas incorporating $\Theta_{CW}$ gives good fits at each temperature. However, the extracted $\Theta_{CW}$, rather than being fixed as would be expected, exhibits a strong temperature dependence, becoming increasingly positive as temperature rises, indicating stronger ferromagnetic interactions. In $V_2O_3$, we expect the interactions between the neighboring V ions to be antiferromagnetic. Thus, between this analysis and the observed systematic thickness dependence in $V_2O_3$, we conclude that the LSSE observed in $V_2O_3$ is not the interfacial paramagnetic SSE.

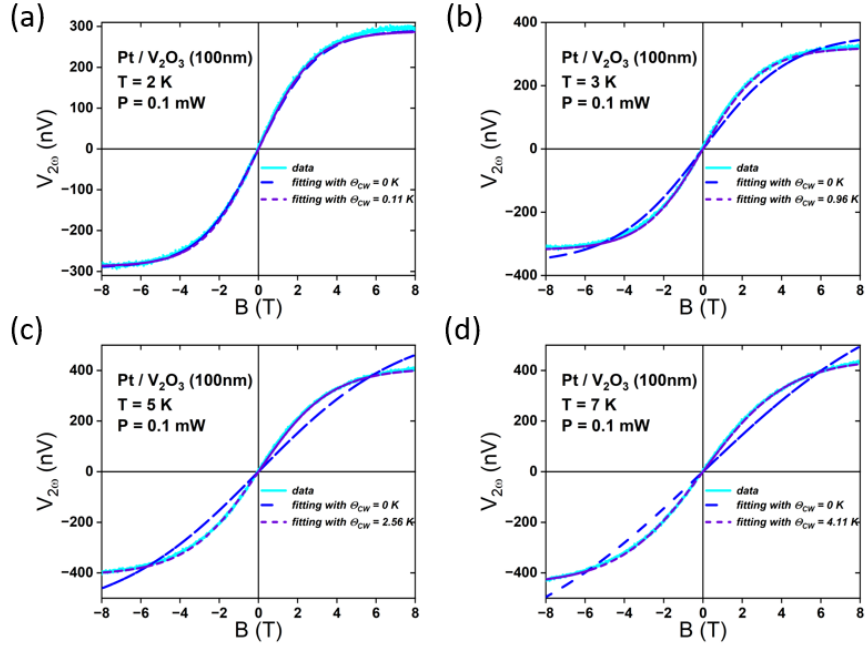

FIG 4. (color online) Comparison of the measured second harmonic voltage and fitting with and without the parameter $\Theta_{CW}$ at (a) 2 K, (b) 3 K, (c) 5 K and (d) 7 K. At each temperature, the blue line is the data points, the cyan dash line is the fitting without $\Theta_{CW}$, and the purple double dash line is the fitting with $\Theta_{CW}$.

In conclusion, we observe a temperature-dependent longitudinal spin Seebeck response in $V_2O_3$ thin films within the low-temperature antiferromagnetic phase. The behavior of the measured LSSE voltage shares a significant feature with the paramagnetic LSSE response observed in $VO_2$.

However, the magnitude of the LSSE signal decreases consistently with increasing thickness, seemingly ruling out a dominant interfacial contribution to the SSE. This negative correlation between the SSE and the thickness suggests that the energy relaxation length is shorter than 50 nm in $V_2O_3$, consistent with strong spin-lattice coupling in $V_2O_3$. Additional studies of magnon modes in $V_2O_3$ are required to resolve the nature of spin transport in this correlated system.

**Acknowledgements**

RL, TJL, GE and DN acknowledge support from NSF DMR-2102028 for spin Seebeck measurements. DN acknowledges DOE BES award DE-FG02-06ER46337 for investigations into nanostructures based on transition metal oxide correlated materials. HN, ACB, EQ and IKS were supported by the Department of Energy's Office of Basic Energy Science, under grant # DE-FG02-87ER45332.

.

*Authors Declarations*

*Conflict of Interest*

The authors have no conflicts to disclose.

*Data Availability*

The data that support these findings are available on Zenodo (https://doi.org/10.5281/zenodo.16990523) [54].